\documentclass{IEEEcsmag}

\usepackage[colorlinks,urlcolor=blue,linkcolor=blue,citecolor=blue]{hyperref}

\usepackage{upmath}
\usepackage{amssymb}

\jvol{XX}
\jnum{XX}
\paper{8}
\jmonth{May/June}
\jname{}
\pubyear{2020}

\begin{document}

\title{Causal Modeling of Twitter Activity During COVID-19}

\author{Oguzhan Gencoglu}
\affil{Tampere University, Faculty of Medicine and Health Technology, Tampere, Finland}

\author{Mathias Gruber}
\affil{{LEO Pharma, Copenhagen, Denmark}}

\begin{abstract}
Understanding the characteristics of public attention and sentiment is an essential prerequisite for appropriate crisis management during adverse health events. This is even more crucial during a pandemic such as COVID-19, as primary responsibility of risk management is not centralized to a single institution, but distributed across society. While numerous studies utilize Twitter data in descriptive or predictive context during COVID-19 pandemic, causal modeling of public attention has not been investigated. In this study, we propose a causal inference approach to discover and quantify causal relationships between pandemic characteristics (e.g. number of infections and deaths) and Twitter activity as well as public sentiment. Our results show that the proposed method can successfully capture the epidemiological domain knowledge and identify variables that affect public attention and sentiment. We believe our work contributes to the field of infodemiology by distinguishing events that correlate with public attention from events that cause public attention.

%to distinguish epidemiological events that correlate with public attention from epidemiological events that cause public attention.

%identify truly causal relationships between pandemic characteristics and public behaviour remains crucial for devising impactful public policies. 

%Distinguishing epidemiological events that correlate with public attention from epidemiological events that cause public attention is crucial for constructing impactful public health policies.

%We hypothesize that daily Twitter activity and sentiment during the COVID-19 pandemic has a causal relationship with the characteristics of the pandemic. We propose a causal inference approach to distinguish epidemiological events that correlate with public attention from epidemiological events that cause public attention.

%We first employ a structure learning method to automatically construct a graphical causal structure in a data-driven manner. Then, we utilize \textit{Bayesian Networks} (BNs) to learn conditional probability distributions of daily Twitter activity (number of daily tweets) and average public sentiment with respect to several pandemic characteristics such as total number of deaths and number of new infections. Our results show that the proposed structure discovery method can successfully capture the epidemiological domain knowledge.
\end{abstract}
\maketitle

\chapterinitial{On} 11 March 2020, Coronavirus disease 2019 (COVID-19) was declared a pandemic by the World Health Organization \cite{cucinotta2020declares} and more than 30 million people have been infected by it as of 19 September 2020 \cite{dong2020interactive}. During such crises, capturing the dissemination of information, monitoring public opinion, observing compliance to measures, preventing disinformation, and relaying timely information is crucial for risk communication and decision-making about public health \cite{van2020using}. Previous national and global adverse health events show that social media surveillance can be utilized successfully for systematic monitoring of public perception in real-time due to its instantaneous global coverage \cite{signorini2011use,ji2013monitoring,ji2015twitter,weeg2015using,mollema2015disease,jordan2019using}.

Due to its large number of users, Twitter has been the primary social media platform for acquiring, sharing, and spreading information during global adverse events, including the COVID-19 pandemic \cite{rosenberg2020twitter}. Especially during the early stages of the COVID-19 pandemic, millions of posts have been tweeted in a span of couple of weeks by users, i.e., citizens, politicians, corporations, and governmental institutions \cite{chen2020covid,gao2020naist,ieee2020data,aguilar2020dataset}. Consequently, numerous studies proposed and utilized Twitter as a data source for extracting insights on public health as well as insights on public attention during the COVID-19 pandemic. Focus of these studies include content analysis \cite{thelwall2020retweeting}, topic modeling \cite{sha2020dynamic}, sentiment analysis \cite{wong2020paradox}, nowcasting or forecasting of the disease \cite{turiel2020wisdom}, early detection of the outbreak \cite{gharavi2020early}, quantifying and detecting misinformation, disinformation, or conspiracies \cite{chary2020geospatial}, and measuring public attitude towards relevant health concepts (e.g. social distancing or working from home) \cite{kayes2020automated}.

Despite such abundance of studies on manual or automatic analysis of social media data during COVID-19, \textit{causal} modeling of relationships between characteristics of the pandemic and social media activity has not been investigated at all, as of September 2020. While descriptive statistical analysis (e.g. correlation, cluster, or exploratory analysis) is beneficial for pattern and hypothesis discovery, and standard machine learning methods are effective in predictive modeling of those patterns, causal inference of relevant phenomena will not be possible without causal computational modeling. Causal modeling in the context of social media and pandemic can enable the optimization of onset of risk communication interventions to increase dissemination of accurate information. Similarly, it can be utilized to prevent acute propagation of negative sentiment with timely interventions. Consequently, such causal modeling can help risk communication policies to shift from alerting people to reassuring them. Furthermore, causal modeling enables simulation of what-if scenarios to enhance disaster preparedness. Therefore, as public decision-making can benefit from adequate assessment of public attention and correct understanding of underlying causes affecting it, we hereby propose causal modeling of Twitter activity.

We hypothesize that daily Twitter activity and sentiment during the COVID-19 pandemic has a causal relationship with the characteristics of the pandemic as well as with certain country statistics. We propose a structural causal modeling approach for discovering causal relationships and quantifying likelihood of events under various conditions (i.e. causal queries). To validate our approach, we collect close to 1 million tweets with location information spanning 57 days and identify several attributes of COVID-19 pandemic that might affect Twitter activity. We first employ a structure learning method to automatically construct a graphical causal structure in a data-driven manner. Then, we utilize \textit{Bayesian Networks} (BNs) to learn conditional probability distributions of daily Twitter activity (number of daily tweets) and average public sentiment with respect to several pandemic characteristics such as total number of deaths and number of new infections. Our results show that the proposed structure discovery method can successfully capture the epidemiological domain knowledge. Furthermore, causal inference of daily Twitter activity with cross-validation across 12 countries show that our approach provides accurate predictions of Twitter activity with interpretable and intuitive results. We release the full source code of our study\footnote{\url{https://github.com/ogencoglu/causal_twitter_modeling_covid19}}. We believe our study contributes to the field of infodemiology by proposing causal modeling of public attention during the crisis of COVID-19 pandemic.

\section{GOING BEYOND CORRELATIONS}
Use of observational data from social media was proven to be beneficial in systematic monitoring of public opinion during adverse health events \cite{signorini2011use,ji2013monitoring,ji2015twitter,weeg2015using,mollema2015disease,jordan2019using}. Such utilization of large, publicly available data becomes even more relevant during a global pandemic such as COVID-19, as neither enough time nor a practical way to run variety of randomized control trials for quantifying public opinion exist. Furthermore, as disease containment measures (e.g. lockdowns, quarantines, and curfews), associated financial issues (e.g. due to inability to work), and changes in social dynamics may impact mental health negatively \cite{wang2020immediate,cullen2020mental,brooks2020psychological}, opinion surveillance methods that do not carry the risk of further stressing of the participants are pertinent.

Themes of previous studies that focus on exploration of, description of, correlation of, or predictive modeling with Twitter data during COVID-19 pandemic include sentiment analysis \cite{wong2020paradox,dubey2020analysing,duong2020ivory,medford2020infodemic,samuel2020covid}, public attitude/interest measurement \cite{kayes2020automated,batooli2020measuring,kwon2020defining,cinelli2020covid}, content analysis \cite{park2020conversations,thelwall2020covid,thelwall2020retweeting,alshaabi2020world,lopez2020understanding,dewhurst2020divergent}, topic modeling \cite{abd2020top,sha2020dynamic,wicke2020framing,jarynowski2020trends,ordun2020exploratory,duong2020ivory,medford2020infodemic}, analysis of misinformation, disinformation, or conspiracies \cite{yang2020prevalence,chary2020geospatial,ahmed2020covid,ferrara2020covid,bridgman2020causes,ahmed2020dangerous,gallotti2020assessing}, outbreak detection or disease nowcasting/forecasting \cite{gharavi2020early,turiel2020wisdom}, and more \cite{golder2020chronological,sarker2020self,li2020we,xu2020twitter,lyu2020sense,schild2020go}. Similarly, data from other social media channels (e.g. Weibo, Reddit, Facebook) or search engine statistics are utilized for parallel analyses related to COVID-19 pandemic as well \cite{rovetta2020covid,shahsavari2020conspiracy,li2020data,li2020impact,velasquez2020hate,zhao2020chinese,li2020characterizing,lampos2020tracking,boberg2020pandemic,jelodar2020deep,liu2020machine,hou2020assessment,stokes2019public,shen2020reports,chen2020unpacking,lucas2020online,pekoz2020covid}. While these studies reveal important information and patterns, they do not attempt to uncover or model causal relationships between the attributes of COVID-19 pandemic and social media activity. As \textit{correlation does not imply causation} (e.g. spurious correlations), the ability to identify truly causal relationships between pandemic characteristics and public behaviour (online or not) remains crucial for devising public policies that are more impactful. Without causal understanding, our efforts and decisions on risk communication, public health engagement, health intervention timing, and adjustment of resources for fighting disinformation, fearmongering, and alarmism will stay subpar.

The task of forging causal models comes with numerous challenges in various domains because, typically, domain knowledge and significant amount of time from the experts is required. For substantially complex phenomena such as a pandemic due to a novel virus, diagnosing causal attributions becomes even harder. Therefore, learning causal relationships automatically from observational data has been studied in machine learning. One of the primary challenges for this pursuit is that numerous latent variables that we can not observe exist in real world problems. In fact, numerous other latent variables that we are not even aware of may exist as well. As latent variables can induce statistical correlations between observed variables that do not have a causal relationship, \textit{confounding factors} arise. While this phenomenon may not exhibit a considerable problem in standard probabilistic models, causal modeling suffers from it immensely.

Several machine learning methods are proposed for learning causal structures from observational data and some allow combination of statistical information (learned from the data) and domain expertise \cite{ellis2008learning,koller2009probabilistic}. Bayesian networks are frequently utilized frameworks for learning models once the causal structure is fixed. As probabilistic graphical models, BNs flexibly unify graphical models, structural equations, and counterfactual logic \cite{rubin2005causal,pearl2010introduction,koller2009probabilistic,pearl2009causality}. A causal BN consists of a directed acyclic graph (DAG) in which nodes correspond to random variables and edges correspond to direct causal influence of one node on another \cite{koller2009probabilistic}. This compact representation of high-dimensional probability spaces (e.g. joint probability distributions) provides intuitive and explainable models for us. In addition, BNs allow not only straightforward observational computations (e.g. calculation of marginal probabilities) but also interventional ones (e.g. \textit{do-calculus}), enabling simulations of various what-if scenarios.

% health informatics and social health.

% As explicit representation of a joint distribution over a set of random variables is computationally unmanageable, use graph structure such as BNs 
% Main goal is to isolate the component of the correlation that is due to the \textbf{causal effect} of X on Y .

\begin{figure*}
\centerline{\includegraphics[width=\textwidth,trim={0 5.4cm 0 0},clip]{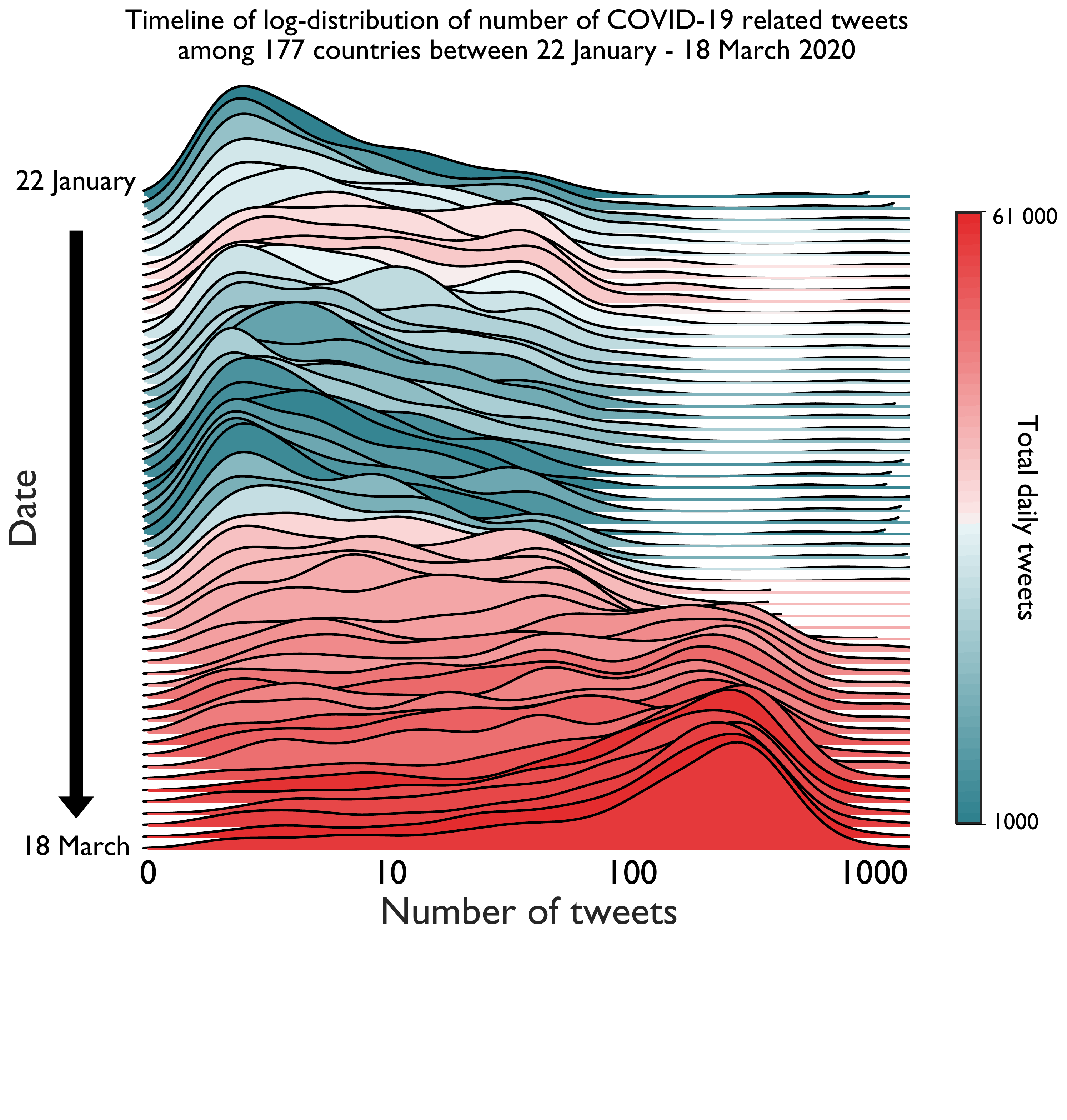}}
\caption{Evolution of COVID-19 related Twitter activity between 22 January - 18 March 2020.}
\label{fig1}
\centering
\end{figure*}

\section{METHODS}
\subsection{Data}
We primarily utilized two data sources for our study, i.e., daily number of officially reported COVID-19 infections and deaths from ``COVID-19 Data Repository'' by the Center for Systems Science and Engineering at Johns Hopkins University \cite{dong2020interactive} and daily count of COVID-19 related tweets from Twitter \cite{twitter}. A 57 day period between 22 January-18 March 2020 is chosen for this study to represent the early stages of the pandemic when disease characteristics are less known and public panic is elevated. We collected 954,902 tweets that have location information from Twitter by searching for \textit{\#covid19} and \textit{\#coronavirus} hashtags. Similar to other studies \cite{turiel2020wisdom,gallotti2020assessing,chary2020geospatial}, geolocation of the tweets is inferred either by using user geo-tagging or geo-coding the information available in users' profiles. Timeline of daily log-distribution of collected tweet counts among 177 countries can be examined from Figure \ref{fig1}. The trend shows an increasing prevalence of high daily number of tweets as the pandemic spreads across the globe with time. % 1,161 and 66,XXX tweets have been observed on 22 January and 18 March, respectively.

\begin{figure*}
\centerline{\includegraphics[width=\textwidth]{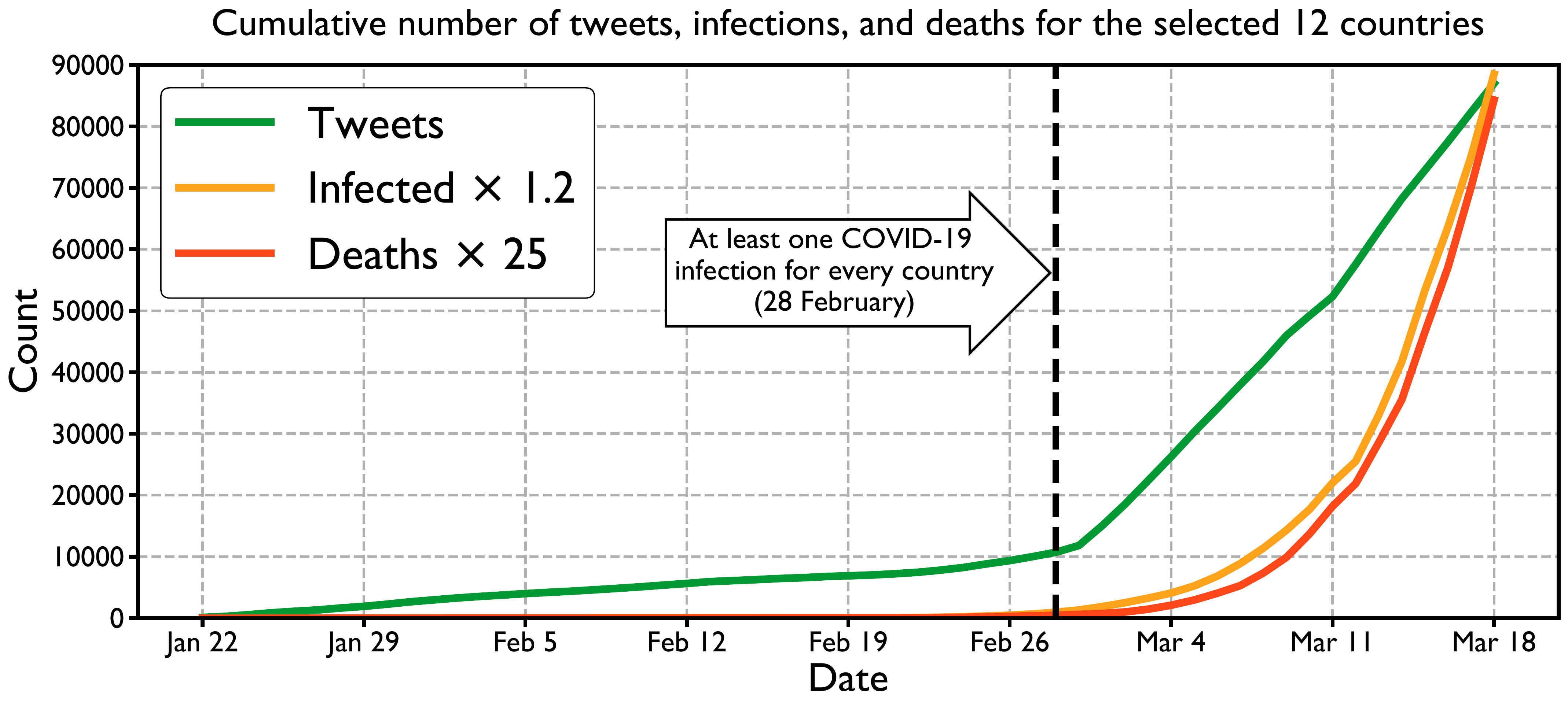}}
\caption{Cumulative counts of Twitter activity and COVID-19 statistics for the selected countries during the study period.}
\label{fig2}
\centering
\end{figure*}

We select the following 12 countries for our causal modeling analysis: Italy, Spain, Germany, France, Switzerland, United Kingdom, Netherlands, Norway, Austria, Belgium, Sweden, and Denmark. These are the countries with substantial number of reported COVID-19 cases (listed in descending order) in Europe as of 18 March 2020, yet still exhibiting a high diversity in terms of the timeline of the pandemic. For instance, while Italy located further in the pandemic timeline due to being hit first in Europe, United Kingdom could be considered in the very initial stages of it for the analysis period of our study. Figure \ref{fig2} depicts the cumulative number of tweet counts alongside with that of reported infections and deaths for the selected countries. Evident correlations between these variables can be noticed. A sharp increase in Twitter activity is observed after 28-29 February, which corresponds to the period of each country having at least one confirmed COVID-19 case.

\subsection{Feature Selection and Engineering}

In order to characterize the pandemic straightforwardly, we calculate the following six features (attributes) from the official COVID-19 incident statistics for each day for 12 selected countries: (1) \textit{total number of infections up to that day} (normalized by the country's population), (2) \textit{number of new infections} (normalized by the country's population), (3) \textit{percentage increase in infections} (with respect to previous day), and the same three statistics for \textit{deaths} (4-5-6).

Recent epidemiological studies on COVID-19 reveal the following: people over the age 65 are the primary risk group both for infection and mortality \cite{dowd2020demographic,guo2020origin,yang2020clinical,wang2020updated} and human-to-human transmission of the virus is largely occurring among family members or among people who co-reside \cite{world2020report,li2020asymptomatic,guo2020origin}. In order to be able to test whether our approach can capture this scientific domain knowledge or not, we collect the following two features for each country: (7) \textit{percentage of population over the age of 65} \cite{worldbank} and (8) \textit{percentage of single-person households} \cite{singlehousehold}. Finally, as we know that popularity of Twitter in a country and announcement of national lockdown (e.g. closing of schools, banning of gatherings) unequivocally affect the Twitter activity in that country, we add (9) \textit{percentage of population using Twitter} \cite{twitterstats} and (10) \textit{is\_lockdown\_announced?} (3 day period is encoded as \verb|Yes| if government restriction is announced \cite{lockdowns}, \verb|No| otherwise) features as well. We represent Twitter activity by simply counting the (11) \textit{number of daily tweets} (normalized by the country's population). We also calculate the (12) \textit{average daily sentiment} (in range [-1, 1]) of English tweets (corresponding to over 80\% of all tweets) by utilizing a pre-trained sentiment classifier (DistilBERT \cite{sanh2019distilbert}). We treat each day as an observation and represent each day with these 12 attributes ($n=12$) for structure learning, resulting in a feature matrix of dimensions $684 \times 12$. 684 observations come from 12 countries times 57 days.

For the purpose of increasing interpretability, we discretize the daily numerical features by mapping them to 2 categorical levels, namely \verb|High| or \verb|Low|. Features related to the pandemic (infections and deaths) and Twitter activity employ a cut-off value of 75th percentile and remaining numerical features employ a cut-off value of 50th percentile (corresponding to median). Such categorization, for instance, turns the numerical value of ``population-normalized increase in deaths of $1.7325 \times 10^{-7}$'' into a relatively calculated category of \verb|High| for a given day. Sentiment scores are mapped to \verb|Positive| ($\geq0$) or \verb|Negative| ($<0$) as well.

\subsection{Structure Learning and Causal Inference}

In structure learning we would like to learn a directed acyclic graph, $G$, that describes the conditional dependencies between variables in a given data matrix. A typical formulation of this problem is a \textit{structural equation model} (more generally a \textit{generalized linear model}) in which a weighted adjacency matrix, $W \in \mathbb{R}^{n \times n}$, defines the graph. This is essentially a parametric model that enables operations on the continuous space of $n \times n$ matrices instead of discrete space of DAGs. Such formulation enables a score-based learning of DAGs, i.e.,
\begin{equation}
\begin{aligned}
\quad\quad\quad \min_{W \in \mathbb{R}^{n \times n}} & L(W)\\
\quad\quad\quad \mbox{subject to} \;\; & G(W) \in \mbox{DAGs}
\end{aligned}
\label{eq1}
\end{equation}
where $G(W)$ is the $n$-node graph induced by the weighted adjacency matrix, $W$, and $L$ is the score/loss function to be minimized. Even though the loss function is continuous, solving Equation \ref{eq1} is still a non-convex, combinatorial optimization problem as the acyclicity constraint is discrete and difficult to enforce. Note that acyclicity is a strict requirement for causal graphs. In order to tackle this problem efficiently, we utilize the recently proposed NOTEARS (corresponding to \textit{Non-combinatorial Optimization via Trace Exponential and Augmented lagRangian for Structure learning}) algorithm for structure learning \cite{zheng2018dags}.

NOTEARS algorithm discovers a directed acyclic graph from the observational data by re-formulating the structure learning problem as a purely continuous optimization. This approach differs significantly from existing work in the field which predominantly operates on discrete space of graphs. Re-formulation is achieved by introducing a continuous measure of ``DAG-ness'', $h(W)$, which quantifies the severity of violations from acyclicity as $W$ changes. Consequently, the problem formulation becomes
\begin{equation}
\begin{aligned}
\quad\quad\quad \min_{W \in \mathbb{R}^{n \times n}} & L(W)\\
\quad\quad\quad \mbox{subject to} \;\; & h(W) = 0
\end{aligned}
\label{eq2}
\end{equation}
which enables utilization of standard numerical solving methods and scales cubically, $\mathcal{O}(n^3)$, with the number of variables instead of exponentially as in other structure learning methods. We have chosen the score to be the least squared loss (can be any smooth loss function) with $l_1$-regularization term to discover a sparse DAG and use a gradient-based minimizer to solve Equation \ref{eq2}. In our context, we discover such an adjacency matrix that the graph it defines encodes the dependencies between our features in a close-to-optimal manner (finding the global optimum is NP-hard \cite{chickering1996learning,chickering2004large}) and is a DAG. Efficiency of this approach enables structure learning in a scalable manner.

As NOTEARS algorithm allows incorporation of expert knowledge, we also put certain constraints on the structure in our experiment. These constraints correspond to prohibited causal attributions based on simple logical assumptions, e.g. Twitter activity on a given day can not have a causal effect on number of deaths from COVID-19 on that day. Once the structure is learned (both by data and logical constraints), we treat it as a causal model and learn the parameters of a Bayesian network on it with the training data in order to capture the conditional dependencies between variables. During inference on test data, probabilities of each possible state of a node with respect to the given input data is computed from the conditional probability distributions.

Our approach allows straightforward querying of the model with varying observations. For instance for a given day, the probability of Twitter activity being \verb|High|, when total number of infections are \verb|Low| and new deaths are \verb|High|, i.e.,
\begin{equation}
\begin{aligned}
& \Pr(\text{Twitter Activity}=\verb|H| \mid &  \\ 
& \;\; \text{Total Infections}=\verb|H|, \; \text{New Deaths}=\verb|L|), &
\end{aligned}
\label{eq3}
\end{equation}
can be computed by propagating the impact of these queries through the nodes of interest. By utilizing this property of our approach, we compute marginal probabilities for gaining further insights on likelihoods of various events.

\begin{figure*}
\centerline{\includegraphics[width=\textwidth,trim={0.6 3.1cm 0.5 0.9},clip]{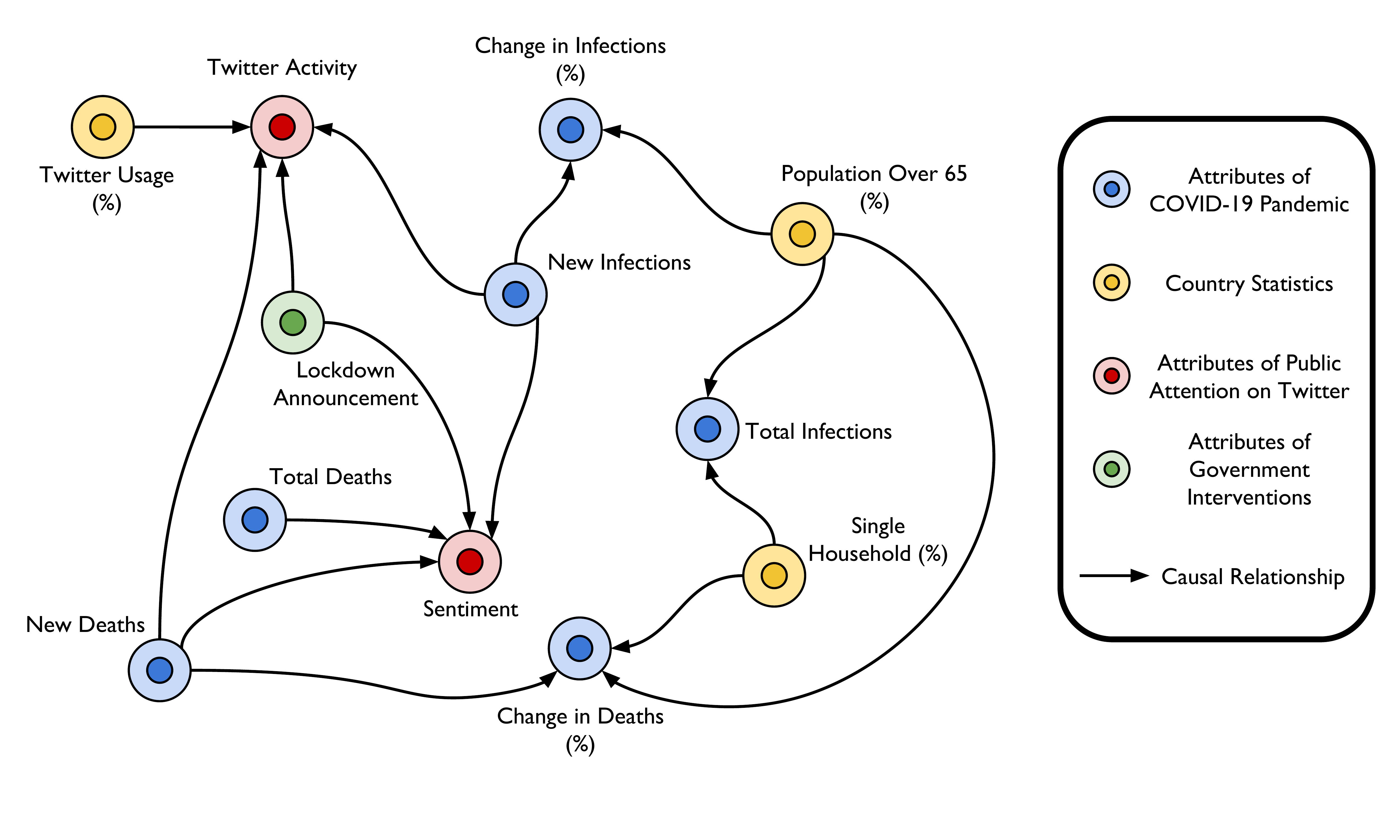}}
\caption{Discovered graph depicting causal relationships between various attributes.}
\label{fig3}
\centering
\end{figure*}

Essentially, we expect two observations from our experiment. First, we expect the structure learning algorithm to discover the causal relations verified by domain/expert knowledge (e.g. \% of single-person households and \% of 65+ people affecting infections) and common sense/elementary algebra (e.g. new deaths affecting percentage change in deaths). Second, we expect the calculated likelihoods from the Bayesian network are in parallel with domain knowledge as well, e.g. high \% of people over 65 increasing the marginal likelihood of deaths instead of decreasing it or high \% of single households (better social isolation) decreasing the marginal likelihood of infections instead of increasing it. Realization of these expectations will show that the proposed method can indeed capture causal relationships and will increase our confidence in discovered relationships between the pandemic attributes and Twitter activity as well as confidence in corresponding likelihoods.

\subsection{Evaluation}

We validate our approach first by inspecting whether the expected causal relationships (e.g. domain knowledge on COVID-19) are captured or not. Then, we infer the Twitter activity of each day from the learned Bayesian Network. Essentially, this corresponds to a binary classification task, i.e., predicting the Twitter activity as \verb|High| or \verb|Low| from the rest of the variables. We utilize a Leave-One-Country-Out (LOCO) cross-validation scheme in which each fold consists of training set from 11 countries (627 samples) and test set (57 samples) from the remaining country. We do not perform standard k-fold cross-validation as we would like to measure the generalization performance across countries and prevent overly optimistic results. Therefore, we ensure that the observations from the same country fall in the same set (either training or test) for every fold. We evaluate the performance of our approach by calculating the average Area Under the Receiver Operating Characteristic curve (AUROC) of the cross-validation runs. For quantifying the causal effect of characteristics of pandemic and relevant country statistics on Twitter activity, we report likelihoods from the model by querying various conditions.

\section{RESULTS}

%The correlation between the number of infections and number of tweets is XX ($p < 10^{-3}$) and between the number of deaths and number of tweets is XX ($p < 10^{-3}$). Cumulative number of tweets, infections and deaths between 22 January-18 March 2020 in Figure 2 shows that .... 

The jointly (with statistical learning from data and user-defined logical constraints) discovered causal model by the structure learning algorithm can be examined from Figure \ref{fig3}. Different families of attributes are colored differently for ease of inspection: blue for COVID-19 pandemic related variables, yellow for country-specific statistics, green for government interventions, and red for representing variables related to public attention and sentiment in Twitter. Daily Twitter activity is affected by 4 variables, namely Twitter usage statistics of that country, new infections on that day, new deaths on that day, and whether national lockdown is announced or not. Similarly, 4 variables affecting the average daily sentiment in Twitter are new infections on that day, new deaths on that day, total deaths up to that day, and again lockdown announcements. Total number of infections did not show any causal effect on Twitter activity or on average public sentiment.

Leave-One-Country-Out cross-validation results in terms of AUROCs can be seen in Table \ref{tab1}. Each row in the table corresponds to a cross-validation fold in which the Twitter activity in that particular country was tried to be predicted. The Bayesian network model achieves an average AUROC score of 0.833 across countries when trying to infer the Twitter activity from the rest of the variables for a given day. Daily Twitter patterns of Germany, Italy, and Sweden show very high predictability with AUROC scores above 0.97. United Kingdom shows the worst predictability with an AUROC of 0.68.

Calculation of marginal probabilities for several queries are presented in Table \ref{tab2}. Public attention and sentiment-related target variables and states are set to \verb|High| Twitter Activity and \verb|Negative| Sentiment.

\section{DISCUSSION}

\begin{table}
\caption{AUC result for each fold of Leave-One-Country-Out cross-validation.}
\small
\hspace{1cm}
\begin{tabular*}{17.5pc}{r|c}
\textbf{Cross Validation Test Country} & \textbf{AUC} \\
\cline{1-2}
Austria & 0.798 \\
Belgium & 0.728 \\
Denmark & 0.831 \\
France & 0.776 \\
Germany & 0.992 \\
Italy & 0.976 \\
Netherlands & 0.746 \\
Norway & 0.907 \\
Spain & 0.766 \\
Sweden & 0.998 \\
Switzerland & 0.789 \\
United Kingdom & 0.684 \\
\cline{1-2}
\textit{Average} & \textit{0.833} \\
\cline{1-2}
\end{tabular*}
\label{tab1}
\end{table}

\begin{table}
\caption{Examples of queries and computed marginal probabilities for Twitter activity and average sentiment.}
\small
\begin{tabular*}{\columnwidth}{c|c|c}
\textbf{Query} & \textbf{Variable and State} & $\boldsymbol{\Pr()}$ \\ 
\hline
Single-p. hh. (\%) $ = $ \verb|H| & Total Infections $ = $ \verb|H| & 0.178 \\
65+ (\%) $ = $ \verb|L| &  & \\
\hline
Single-p. hh. (\%) $ = $ \verb|L| & Total Infections $ = $ \verb|H| & 0.241 \\
65+ (\%) $ = $ \verb|H| &  & \\
\hline \hline 
New Infections $ = $ \verb|H| & Twitter Activity $ = $ \verb|H| & 0.496 \\
New Deaths $ = $ \verb|H| &  & \\
\hline
New Infections $ = $ \verb|L| & Twitter Activity $ = $ \verb|H| & 0.184 \\
New Deaths $ = $ \verb|L| &  & \\
\hline \hline
New Infections $ = $ \verb|H| &  &  \\
New Deaths $ = $ \verb|H| & Twitter Activity $ = $ \verb|H|  & 0.800 \\
Twitter Usage $ = $ \verb|H| &  & \\
Lockdown Ann. $ = $ \verb|Yes| &  & \\
\hline
New Infections $ = $ \verb|L| &  &  \\
New Deaths $ = $ \verb|L| & Twitter Activity $ = $ \verb|H| & 0.120 \\
Twitter Usage $ = $ \verb|L| &  & \\
Lockdown Ann. $ = $ \verb|No| &  & \\
\hline \hline
New Deaths $ = $ \verb|H| & Sentiment $ = $ \verb|Neg| & 0.624 \\
\hline
New Deaths $ = $ \verb|L| & Sentiment $ = $ \verb|Neg| & 0.277 \\
\hline \hline 
Total Deaths $ = $ \verb|H| & Sentiment $ = $ \verb|Neg| & 0.344 \\
\hline
Total Deaths $ = $ \verb|L| & Sentiment $ = $ \verb|Neg| & 0.290 \\
\hline \hline
Lockdown Ann. $ = $ \verb|Yes| & Sentiment $ = $ \verb|Neg| & 0.501 \\
\hline
Lockdown Ann. $ = $ \verb|No| & Sentiment $ = $ \verb|Neg| & 0.286 \\
\hline
\end{tabular*}
\label{tab2}
\end{table}

% Correlations only: Analyzed Twitter regarding online collective attention to the COVID-19 pandemic~\cite{dewhurst2020divergent}.

By analyzing observational data, we attempt to discover causal associations between national COVID-19 patterns and Twitter activity as well as public sentiment during the early stages of the pandemic. Some of our findings are expected associations such as popularity of Twitter in a country (Twitter usage) affecting Twitter activity. Other expected causal relationships were new deaths affecting change in deaths and new infections affecting change in infections, due to trivial mathematical definitions. These were captured successfully as well. It is important to note that no causal relationship between infection statistics and death statistics was discovered which might seem against intuition. This is because in this study we treat each day as an observation in our modeling and do not create time-lagged version of variables. While some of our results imply expected associations, we also observe more interesting implications that are in alignment with recent scientific literature on COVID-19. For instance, percentage of single-person households affects the total number of COVID-19 infections. Similarly, percentage of 65+ population affects the percentage change in deaths (essentially corresponding to rate of deaths). When the queries regarding domain knowledge are examined, we see that low percentage of single-person households (less social isolation) and high percentage of 65+ population increases the probability of total infections being high when compared to the opposite settings. This is in line with recent scientific literature on COVID-19 transmission characteristics \cite{dowd2020demographic,world2020report,li2020asymptomatic,guo2020origin,yang2020clinical,wang2020updated}.

By inferring Twitter activity, we show the generalization ability of causal inference across 12 countries with reasonable accuracy. Factors affecting Twitter activity and sentiment are discussion-worthy as well. By observing correlations, Wong et al. hints that there may be a link between announcement of new infections and Twitter activity \cite{wong2020paradox}. Our results in Figure \ref{fig3} and Table \ref{tab2} suggest the same with a causal point of view. Similarly, our finding of negative impact of declaration of government measures on public sentiment is also in parallel with recent research. By analyzing Chinese social media, Li et al. show that official declaration of COVID-19 (epidemic at that time) correlates with increased negative emotions such as anxiety, depression, and indignation \cite{li2020impact}. When new infections, new deaths, total deaths are high and an announcement of lockdown is made, Twitter activity on that day becomes more than 6 times more likely than when the situation is opposite (probabilities of 0.8 vs. 0.12). High number of new deaths for a given day causes the sentiment to be much more negative than low number of new deaths (probabilities of 0.624 vs. 0.277). Similarly, an announcement of lockdown is causally associated with an increase in negative sentiment in Twitter (probabilities of 0.501 vs. 0.286).

As it is important to observe the countries that are ahead in terms of pandemic timeline and learn the behaviour of the pandemic, it is equally important to understand also the public attention and sentiment characteristics from those countries. Wise et al. show that risk perception of people and their frequency of engagement in protective behaviour change during the early stages of the pandemic \cite{wise2020changes}. Inference of such patterns in a causal manner from social media can aid us in the pursuit of timely decisions and suitable policy-making, and consequently, high public engagement. After all, primary responsibility of risk management during a global pandemic is not centralized to a single institution, but distributed across society. For example, Zhong et al. shows that people's adherence to COVID-19 control measures is affected by their knowledge and attitudes towards it \cite{zhong2020knowledge}. In that regard, computational methods such as causal inference and causal reasoning can help us disentangle correlations and causation between the observed variables of the adverse phenomenon.

In real-world scenarios, it is virtually impossible to correctly identify all the causal associations due to presence of numerous confounding factors. As in with all methods in machine learning, a trade-off between false positive associations and false negative ones exists in our approach as well. While we rely on official COVID-19 statistics, testing and reporting methodologies as well as policies can change during the course of the pandemic. Furthermore, in the context of this study, ground truth causal associations do not exist even for a few variables, preventing the direct measurement of performance of causal discovery methods. We would like to emphasize that we acknowledge these and other relevant limitations of our study. Our study has further limitations regarding the simplifications on our problem formulation and data. For instance, we do not attempt to model temporal causal relationships in this study, e.g., high deaths numbers having an impact on the public sentiment possibly for several following days. We have not taken into account remarks by famous politicians, public figures, or celebrities which may indeed impact social media discussions. We have not incorporated ``retweets'' or ``likes'' into our models either. We would also like emphasize that with this study we wanted to introduce an uncomplicated example of causal modeling perspective to social media analysis during COVID-19.

Future work includes investigating the effect of dynamics of the pandemic on the spreading mechanisms of information, including relevant health topics in Twitter and other social media. As social media can be exploited for deliberately creating panic and confusion \cite{merchant2020social}, causal inference on patterns of misinformation and disinformation propagation in Twitter will be studied as well. Finally, country-specific models with more granular statistics of the country and time-delayed variables will be investigated for a longer analysis period.

% Twitter users were shown to perform poorly in rumor detection and rush to spread rumors during disasters~\cite{wang2018rumor}.

% Unobserved confounding variables that causes both the dependent and independent variable, resulting in a spurious correlation.

% improving stability of citizens' feelings and urgently prepare clinical practitioners to deliver corresponding therapy foundations for the risk groups and affected people. Getting accustomed to high number of infections and deaths may lead to people breaking rules.

\section{CONCLUSION}

%Quantifying the characteristics of the public attention is an essential prerequisite for appropriate crisis management during severe events such as pandemics. 

Distinguishing epidemiological events that correlate with public attention from epidemiological events that cause public attention is crucial for constructing impactful public health policies. Similarly, monitoring fluctuations of public opinion becomes actionable only if causal relationships are identified. We hope our study serves as a first example of causal inference on social media data for increasing our understanding of factors affecting public attention and sentiment during COVID-19 pandemic.

%adequate education of citizens by the stakeholders.

\bibliographystyle{IEEEtran}
\bibliography{references}

\begin{IEEEbiography}{Oguzhan Gencoglu}{\,}is with the Faculty of Medicine and Health Technology, Tampere University, Finland. His research interests include machine learning and health informatics. Contact him at \url{oguzhan.gencoglu@tuni.fi} or at \url{oguzhangencoglu90@gmail.com}.
\end{IEEEbiography}

\begin{IEEEbiography}{Mathias Gruber}{\,}has an academic background in the natural sciences and computational biotechnology. Current research interests include machine learning applied within the pharmaceutical industry. Can be reached at \url{nano.mathias@gmail.com}.
\end{IEEEbiography}

\end{document}